\documentstyle[12pt,twoside]{article}

\def\a{\alpha}
\def\b{\beta}

\def\d{\delta}
\def\e{\epsilon}                % Also, \varepsilon
\def\f{\phi}                    %       \varphi
\def\g{\gamma}

\def\k{\kappa}
\def\l{\lambda}
\def\m{\mu}
\def\n{\nu}
\def\o{\omega}
\def\p{\pi}                     % Also, \varpi
\def\r{\rho}                    %       \varrho
\def\s{\sigma}                  %       \varsigma
\def\t{\tau}

\def\x{\xi}

\def\P{\Pi}

\def\cl{{\cal L}}

\def\ct{{\cal T}}

\catcode`@=11

\def\un#1{\relax\ifmmode\@@underline#1\else $\@@underline{\hbox{#1}}$\relax\fi}

{\catcode`\'=\active \gdef'{{}~\bgroup\prim@s}}

\def\magstep#1{\ifcase#1 \@m\or 1200\or 1440\or 1728\or 2074\or 2488\or
        2986\fi\relax}

\font\twfvmi=cmmi10\@magscale5
    \skewchar\twfvmi='177
\font\twfvsy=cmsy10\@magscale5
    \skewchar\twfvsy='60
\font\twfvly=lasy10\@magscale5
\font\thtyrm=cmr10\@magscale6

\def\vpt{\textfont\z@\fivrm
  \scriptfont\z@\fivrm \scriptscriptfont\z@\fivrm
\textfont\@ne\fivmi \scriptfont\@ne\fivmi \scriptscriptfont\@ne\fivmi
\textfont\tw@\fivsy \scriptfont\tw@\fivsy \scriptscriptfont\tw@\fivsy
\textfont\thr@@\tenex \scriptfont\thr@@\tenex \scriptscriptfont\thr@@\tenex
\def\prm{\fam\z@\fivrm}%
\def\unboldmath{\everymath{}\everydisplay{}\@nomath
  \unboldmath\fam\@ne\@boldfalse}\@boldfalse
\def\boldmath{\@subfont\boldmath\unboldmath}%
\def\pit{\@getfont\pit\itfam\@vpt{cmti5}}%
\def\psl{\@subfont\sl\it}%
\def\pbf{\@getfont\pbf\bffam\@vpt{cmbx5}}%
\def\ptt{\@subfont\tt\rm}%
\def\psf{\@subfont\sf\rm}%
\def\psc{\@subfont\sc\rm}%
\def\ly{\fam\lyfam\fivly}\textfont\lyfam\fivly
    \scriptfont\lyfam\fivly \scriptscriptfont\lyfam\fivly
\@setstrut\rm}

\def\@vpt{}

\def\vipt{\textfont\z@\sixrm
  \scriptfont\z@\sixrm \scriptscriptfont\z@\sixrm
\textfont\@ne\sixmi \scriptfont\@ne\sixmi \scriptscriptfont\@ne\sixmi
\textfont\tw@\sixsy \scriptfont\tw@\sixsy \scriptscriptfont\tw@\sixsy
\textfont\thr@@\tenex \scriptfont\thr@@\tenex \scriptscriptfont\thr@@\tenex
\def\prm{\fam\z@\sixrm}%
\def\unboldmath{\everymath{}\everydisplay{}\@nomath
  \unboldmath\@boldfalse}\@boldfalse
\def\boldmath{\@subfont\boldmath\unboldmath}%
\def\pit{\@subfont\it\rm}%
\def\psl{\@subfont\sl\rm}%
\def\pbf{\@getfont\pbf\bffam\@vipt{cmbx6}}%
\def\ptt{\@subfont\tt\rm}%
\def\psf{\@subfont\sf\rm}%
\def\psc{\@subfont\sc\rm}%
\def\ly{\fam\lyfam\sixly}\textfont\lyfam\sixly
    \scriptfont\lyfam\sixly \scriptscriptfont\lyfam\sixly
\@setstrut\rm}

\def\@vipt{}

\def\xxxpt{\textfont\z@\thtyrm
  \scriptfont\z@\twfvrm \scriptscriptfont\z@\twtyrm
\textfont\@ne\twfvmi \scriptfont\@ne\twfvmi \scriptscriptfont\@ne\twtymi
\textfont\tw@\twfvsy \scriptfont\tw@\twfvsy \scriptscriptfont\tw@\twtysy
\textfont\thr@@\tenex \scriptfont\thr@@\tenex \scriptscriptfont\thr@@\tenex
\def\unboldmath{\everymath{}\everydisplay{}\@nomath\unboldmath
        \textfont\@ne\twfvmi \textfont\tw@\twfvsy \textfont\lyfam\twfvly
        \@boldfalse}\@boldfalse
\def\boldmath{\@subfont\boldmath\unboldmath}%
\def\prm{\fam\z@\thtyrm}%
\def\pit{\@subfont\it\rm}%
\def\psl{\@subfont\sl\rm}%
\def\pbf{\@getfont\pbf\bffam\@xxxpt{cmbx10\@magscale6}}%
\def\ptt{\@subfont\tt\rm}%
\def\psf{\@subfont\sf\rm}%
\def\psc{\@subfont\sc\rm}%
\def\ly{\fam\lyfam\twfvly}\textfont\lyfam\twfvly
   \scriptfont\lyfam\twfvly \scriptscriptfont\lyfam\twtyly
\@setstrut \rm}

\def\@xxxpt{}

\def\Huge{\@setsize\Huge{36pt}\xxxpt\@xxxpt}

\font\thtymi=cmmi10\@magscale6
    \skewchar\thtymi='177
\font\thtysy=cmsy10\@magscale6
    \skewchar\thtysy='60
\font\thtyly=lasy10\@magscale6
\font\thsirm=cmr12\@magscale6

\def\xxxvipt{\textfont\z@\thsirm
  \scriptfont\z@\thtyrm \scriptscriptfont\z@\twfvrm
\textfont\@ne\thtymi \scriptfont\@ne\thtymi \scriptscriptfont\@ne\twfvmi
\textfont\tw@\thtysy \scriptfont\tw@\thtysy \scriptscriptfont\tw@\twfvsy
\textfont\thr@@\tenex \scriptfont\thr@@\tenex \scriptscriptfont\thr@@\tenex
\def\unboldmath{\everymath{}\everydisplay{}\@nomath\unboldmath
        \textfont\@ne\thtymi \textfont\tw@\thtysy \textfont\lyfam\thtyly
        \@boldfalse}\@boldfalse
\def\boldmath{\@subfont\boldmath\unboldmath}%
\def\prm{\fam\z@\thsirm}%
\def\pit{\@subfont\it\rm}%
\def\psl{\@subfont\sl\rm}%
\def\pbf{\@getfont\pbf\bffam\@xxxpt{cmss12\@magscale6}}%
\def\ptt{\@subfont\tt\rm}%
\def\psf{\@subfont\sf\rm}%
\def\psc{\@subfont\sc\rm}%
\def\ly{\fam\lyfam\thtyly}\textfont\lyfam\thtyly
   \scriptfont\lyfam\thtyly \scriptscriptfont\lyfam\twfvly
\@setstrut \rm}

\def\@xxxvipt{}

\def\HUGE{\@setsize\HUGE{43pt}\xxxvipt\@xxxvipt}

\catcode`@=12

\font\tenex=cmex10 scaled 1200
\def\Sc#1{\hbox{\sc #1}}        % script "
\def\bo{{\raise.05ex\hbox{\large$\Box$}\:}}             % D'Alembertian
\def\cbo{{\,\raise-.15ex\Sc [\,}}                       % curly "
\def\pa{\partial}                                       % curly d
\def\su{\sum}                                           % summation
\def\TH{{\raise.2ex\hbox{$\displaystyle \bigodot$}\mskip-4.7mu \llap H \;}}
\def\face{\hbox{\normalsize$\;\;\:{\raise.9ex\hbox{\oo n}\mskip-13mu \llap
        {${\buildrel{\hbox{\frtnrm ..}}\over\smile}$}}\:$}}     % happy face
\def\Face{{\raise.2ex\hbox{$\displaystyle \bigodot$}\mskip-2.2mu \llap {$\ddot
        \smile$}}}                                      % another "
\def\Lhat{{\bf\rlap{\kern-.09em$\hat{\phantom L}$}L}}
\def\Lcheck{{\bf\rlap{\kern-.09em$\check{\phantom L}$}L}}
\def\sp#1{{}^{#1}}                              % superscript (unaligned)
\def\sb#1{{}_{#1}}                              % sub"
\def\sl#1{\rlap{\hbox{$\mskip 1 mu /$}}#1}      % good slash for lower case
\def\leftrightarrowfill{$\mathsurround=0pt \mathord\leftarrow \mkern-6mu
        \cleaders\hbox{$\mkern-2mu \mathord- \mkern-2mu$}\hfill
        \mkern-6mu \mathord\rightarrow$}
\def\dvec#1{\vbox{\ialign{##\crcr
        \leftrightarrowfill\crcr\noalign{\kern-1pt\nointerlineskip}
        $\hfil\displaystyle{#1}\hfil$\crcr}}}           % <--> accent
\def\dt#1{{\buildrel {\hbox{\LARGE .}} \over {#1}}}     % dot-over for sp/sb
\def\ddt#1{{\buildrel {\hbox{\LARGE .\kern-2pt.}} \over {#1}}}% double dot-over
\def\der#1{{\pa \over \pa {#1}}}                % partial derivative
\def\frac#1#2{{\textstyle{#1\over\vphantom2\smash{\raise.20ex
        \hbox{$\scriptstyle{#2}$}}}}}                   % fraction
\def\ha{\frac12}                                        % 1/2
\def\sfrac#1#2{{\vphantom1\smash{\lower.5ex\hbox{\small$#1$}}\over
        \vphantom1\smash{\raise.4ex\hbox{\small$#2$}}}} % alternate fraction
\def\bfrac#1#2{{\vphantom1\smash{\lower.5ex\hbox{$#1$}}\over
        \vphantom1\smash{\raise.3ex\hbox{$#2$}}}}       % "
\def\afrac#1#2{{\vphantom1\smash{\lower.5ex\hbox{$#1$}}\over#2}}    % "
\def\boxes#1{
        \newcount\num
        \num=1
        \newdimen\downsy
        \downsy=-1.64ex
        \mskip-7.8mu
        \bo
        \loop
        \ifnum\num<#1
        \llap{\raise\num\downsy\hbox{$\bo$}}
        \advance\num by1
        \repeat}
\def\boxup#1#2{\newcount\numup
        \numup=#1
        \advance\numup by-1
        \newdimen\upsy
        \upsy=.82ex
        \mskip7.8mu
        \raise\numup\upsy\hbox{$#2$}}

\newskip\humongous \humongous=0pt plus 1000pt minus 1000pt
\def\caja{\mathsurround=0pt}

\newif\ifdtup
\def\panorama{\global\dtuptrue \openup2\jot \caja
        \everycr{\noalign{\ifdtup \global\dtupfalse
        \vskip-\lineskiplimit \vskip\normallineskiplimit
        \else \penalty\interdisplaylinepenalty \fi}}}
\def\li#1{\panorama \tabskip=\humongous                         % eqalignno
        \halign to\displaywidth{\hfil$\displaystyle{##}$
        \tabskip=0pt&$\displaystyle{{}##}$\hfil
        \tabskip=\humongous&\llap{$##$}\tabskip=0pt
        \crcr#1\crcr}}

\def\CMP{Commun. Math. Phys.}
\def\NP{Nucl. Phys. B}
\def\PL{Phys. Lett. }

\def\ref#1{$\sp{#1]}$}

\parskip=\medskipamount                 % space between paragraphs (TeX)
\def\baselinestretch{1.2}       % magnification for line spacing (LaTeX)
\thicklines                         % thick straight lines for pictures (LaTeX)
\def\title#1#2#3#4{
\begin{document}
        {\hbox to\hsize{#4 \hfill IF- #3}}\par
        \begin{center}\vskip.5in minus.1in {\Large\bf #1}\\[.5in minus.2in]{#2}
        \vskip1.4in minus1.2in {\bf ABSTRACT}\\[.1in]\end{center}
        \begin{quotation}\par}
\def\author#1#2{#1\\[.1in]{\it #2}\\[.1in]}

\def\AMIC{Aleksandar Mikovi\'c\,
\footnote{E-mail address: A.MIKOVIC@IC.AC.UK}
\\[.1in] {\it Theoretical Physics Group, Blackett Laboratory, Imperial
College,\\ Prince Consort Road, London SW7 2BZ, U.K.}\\[.1in]}

\def\WS{W. Siegel\\[.1in] {\it Institute for Theoretical
        Physics,\\ State University of New York, Stony Brook, NY 11794-3840}
        \\[.1in]}

\def\AM{Aleksandar Mikovi\'c\,
\footnote{E-mail address: mikovic@castor.phy.bg.ac.yu}
\\[.1in] {\it Institute of Physics,
 P.O.Box 57, Belgrade 11001, Yugoslavia}\\[.1in]}

\def\AMsazda{Aleksandar Mikovi\'c
\footnote{E-mail address: mikovic@castor.phy.bg.ac.yu}
and Branislav Sazdovi\'c \footnote{E-mail address:
sazdovic@castor.phy.bg.ac.yu}
\\[.1in] {\it Institute of Physics,
 P.O.Box 57, Belgrade 11001, Yugoslavia}\\[.1in]}

\def\endtitle{\par\end{quotation}\vskip3.5in minus2.3in\newpage}

% A4

\def\endabstract{\par\end{quotation}
        \renewcommand{\baselinestretch}{1.2}\small\normalsize}

% Letter

\def\xpar{\par}                                         % \par in loops
\def\letterhead{
        %\phantom m\vskip-.87in
        %\begin{picture}(24,16)(-75,0)
        %\umcp
        %\par
        \centerline{\large\sf IMPERIAL COLLEGE}
        \centerline{\sf Blackett Laboratory}
        \vskip-.07in
        \centerline{\sf Prince Consort Road, SW7 2BZ}
        \rightline{\scriptsize\sf Dr. Aleksandar Mikovi\'c}
        \vskip-.07in
        \rightline{\scriptsize\sf Tel: 071-589-5111/6983}
        \vskip-.07in
        \rightline{\scriptsize\sf E-mail: A.MIKOVIC@IC.AC.UK}}
\def\sig#1{{\leftskip=3.75in\parindent=0in\goodbreak\bigskip{Sincerely yours,}
\nobreak\vskip .7in{#1}\par}}

% Referee report

\def\ree#1#2#3{
        \hfuzz=35pt\hsize=5.5in\textwidth=5.5in
        \begin{document}
        \ttraggedright
        \par
        \noindent Referee report on Manuscript \##1\\
        Title: #2\\
        Authors: #3}

% Book

\def\start#1{\pagestyle{myheadings}\begin{document}\thispagestyle{myheadings}
        \setcounter{page}{#1}}

% Page and section headings and reference stuff

\catcode`@=11

\def\ps@myheadings{\def\@oddhead{\hbox{}\footnotesize\bf\rightmark \hfil
        \thepage}\def\@oddfoot{}\def\@evenhead{\footnotesize\bf
        \thepage\hfil\leftmark\hbox{}}\def\@evenfoot{}
        \def\sectionmark##1{}\def\subsectionmark##1{}
        \topmargin=-.35in\headheight=.17in\headsep=.35in}
\def\ps@acidheadings{\def\@oddhead{\hbox{}\rightmark\hbox{}}
        \def\@oddfoot{\rm\hfil\thepage\hfil}
        \def\@evenhead{\hbox{}\leftmark\hbox{}}\let\@evenfoot\@oddfoot
        \def\sectionmark##1{}\def\subsectionmark##1{}
        \topmargin=-.35in\headheight=.17in\headsep=.35in}

\catcode`@=12

\def\sect#1{\bigskip\medskip\goodbreak\noindent{\large\bf{#1}}\par\nobreak
        \medskip\markright{#1}}
\def\chsc#1#2{\phantom m\vskip.5in\noindent{\LARGE\bf{#1}}\par\vskip.75in
        \noindent{\large\bf{#2}}\par\medskip\markboth{#1}{#2}}
\def\Chsc#1#2#3#4{\phantom m\vskip.5in\noindent\halign{\LARGE\bf##&
        \LARGE\bf##\hfil\cr{#1}&{#2}\cr\noalign{\vskip8pt}&{#3}\cr}\par\vskip
        .75in\noindent{\large\bf{#4}}\par\medskip\markboth{{#1}{#2}{#3}}{#4}}
\def\chap#1{\phantom m\vskip.5in\noindent{\LARGE\bf{#1}}\par\vskip.75in
        \markboth{#1}{#1}}
\def\refs{\bigskip\medskip\goodbreak\noindent{\large\bf{REFERENCES}}\par
        \nobreak\bigskip\markboth{REFERENCES}{REFERENCES}
        \frenchspacing \parskip=0pt \renewcommand{\baselinestretch}{1}\small}
\def\unrefs{\normalsize \nonfrenchspacing \parskip=medskipamount}
\def\Item{\par\hang\textindent}
\def\Itemitem{\par\indent \hangindent2\parindent \textindent}
\def\makelabel#1{\hfil #1}
\def\topic{\par\noindent \hangafter1 \hangindent20pt}
\def\Topic{\par\noindent \hangafter1 \hangindent60pt}

\title{W-Strings on Group Manifolds}
{\AMsazda}{16/94}{December 1994}
We present a procedure for
constructing actions describing propagation of W-strings on group manifolds
by using the Hamiltonian canonical formalism and representations of
W-algebras in terms of Kac-Moody currents. An explicit construction
is given in the case of the $W_3$ string.

\endtitle

W-string (or W-gravity)
theories are higher spin generalizations of ordinary string theories,
such that two-dimensional (2d)
matter is not only coupled to 2d metric but also to
a set of higher spin 2d gauge fields (for a review see [1]).
Since ordinary string theory can be considered as a gauge
theory based on the
Virasoro algebra, one can analogously define a W-string theory as a gauge
theory based on a W-algebra [2] (or any other higher spin conformally extended
algebra [1]).
Actions for a large class of W-string theories have been constructed so far
[3-10]. These actions essentially describe a W-string propagating on a flat
background. In this letter we would like to address the problem of
constructing the action for a W-string propagating on a curved background
by studying the special case of a group manifold.

We are going to use a
general method for constructing gauge invariant actions, based on the
Hamiltonian canonical formalism [9]. This method works if one
knows a representation of the algebra of gauge symmetries in terms of the
coordinates and canonically conjugate momenta. The basic idea is simple:
given a set of canonical pairs $(p\sb i  , q\sp i )$
together with the Hamiltonian $H\sb 0 (p,q)$ and constraints
$G\sb \a (p,q)$ such that
$$ \{G\sb \a, G\sb \b \} = f\sb{\a\b}\sp \g G\sb \g \quad,\eqno(1)$$
$$ \{G\sb \a, H\sb 0 \} = h\sb{\a}\sp \b G\sb \b \quad,\eqno(2)$$
where $\{,\}$ is the Poisson bracket and (1) is the desired algebra
of gauge symmetries, then
the corresponding action is given by
$$ S = \int dt \left( p\sb i \dt{q}\sp i - H\sb 0 - \l\sp \a G\sb \a
\right) \quad.\eqno(3)$$
The parameter $t$ is the time and dot denotes time derivative.
The Lagrange multipliers $\l\sp \a (t)$ play the role of the
gauge fields associated with the gauge symmetries generated by $G_\a$.
The indices $i,\a$ can take
both the discrete and the continious values.
Note that the coeficients $f\sb{\a\b}\sp{\g}$ and $h\sb{\a}\sp{\b}$ can
be arbitrary functions of $p_i$ and $q^i$, and hence the algebra (1) is general
enough to accomodate the case of the $W$ algebras, where the right-hand
side of the Eq. (1) is a non-linear function of the generators.
The action $S$ is invariant under the following gauge transformations
$$ \li{ \d p\sb i = &\e\sp \a \{ G\sb \a , p\sb i \} \cr
        \d q\sp i = &\e\sp \a \{ G\sb \a , q\sp i \} \cr
        \d \l\sp \a = &\dt{\e}\sp \a - \l\sp \b \e\sp \g f\sb{\g\b}\sp \a
                      - \e\sp \b h\sb \b\sp \a \quad.&(4)\cr}$$
It is clear from the the transformation law for
$\l\sp \a$ why they can be identified as gauge fields.

Since we want to describe propagation of a bosonic W-string on a curved
background, the canonical coordinates will be a set of 2d
scalar fields $\f\sp a (\s,\t)$, $a=1,...,n$, where $\s$ is the string
coordinate ($0\le \s \le 2\p$) and $\t$ is the evolution parameter.
$\f^a$ are coordinates on an $n$-dimensional space-time manifold $M$,
and we are going to study the special case when $M$ is a Lie group $G$.
Let $\p\sb a (\s,\t)$ be the canonically conjugate momenta, satisfying
$$ \{ \f\sp a (\s_1 ,\t), \p\sb b (\s_2 ,\t)\} =
\d^{a}_{b}\d (\s_1  - \s_2 ) \quad.
\eqno(5)$$
In order to construct the desired action, we need a canonical representation of
the corresponding W-algebra.
This can be obtained from the canonical analysis of the
Wess-Zumino-Novikov-Witten (WZNW) action and the
fact that the generators of a W-algebra can be obtained as
traces of products of the Kac-Moody currents [7,11,12].

The WZNW action can be written as
$$S\sb 2 = \k\int d^2 \s \left( -\ha\sqrt{-g}g\sp{\m\n} H_{ab} (\f)
+ \e^{\m\n}\ct_{ab}(\f)\right)\pa_\m \f\sp a \pa\sb \n\f\sp b
\quad,\eqno(6)$$
where $g_{\m\n}$ is a 2d metric, $\e^{\m\n}$ is an antisymmetric 2d
tensor density, $\pa_\m = (\pa_0 ,\pa_1 ) =  (\pa_\t , \pa_\s )$,
$H_{ab}$ is the metric on the manifold $G$, while $\ct_{ab}$ is an
antisymmetric
tensor field. These tensors can be defined through left/right invariant
Maurer-Cartan one-forms on $G$
$$ v_+ = g^{-1} dg \quad,\quad v_- =g dg^{-1}= -dg g^{-1}\quad, \quad g \in G
\eqno(7) $$
such that
$$ H_{ab} = Tr (E_{Aa} , E_{Ab} ) = \g_{\a\b}E_{Aa}^\a  E_{Ab}^\b \quad,
\quad v_A = d\f^a E_{Aa}^\a \, t_\a \quad,\quad  \eqno(8)$$
and
$$ Tr(v_+^3) = -6 d\ct \quad,\quad \ct = \ha \ct_{ab}\, d\f^a \wedge d\f^b
\quad,\eqno(9) $$
where $A=\pm$, $E$'s are veilbeins on $G$, $t_\a$ are the generators of
the Lie algebra of $G$, $[t_\a,t_\b] = f\sb{\a\b}\sp{\g} t_\g$, and
$\g_{\a\b} = f\sb{\a\g}\sp{\d} f\sb{\b\d}\sp{\g}$ is the group metric.

The canonical form of the action (6) can be written as
$$ S\sb 2 = \int_{\t_1}^{\t_2} d\t \int_{0}^{2\p}d\s \left( \p\sb a
\dt{\f}\sp a - h\sp{A} T\sb{A}  \right)\quad,\eqno(10)$$
where the constraints $T_A$ are given by
$$ T_A = {1\over 4\k} Tr (J_A^2) ={1\over 4\k} \g^{\a\b} J_{A\a} J_{A\b}
\quad,\eqno(11) $$
$$J_{A\a}= - E_{A\a}^a (\p_a + 2\k \ct_{ab}\f^{\prime b} ) -
(-1)^A \k E_{Aa \a} \f^{\prime a}
\quad,\eqno(12)$$
where primes stand for the $\s$ derivatives. The constraints $T_A$ are $++$ and
$--$ components of the energy-momentum tensor ($A^\pm = A^0 \pm A^1$), and
$T_A$ satisfy the Virasoro algebra
$$ \{T\sb \pm (\s_1),T\sb \pm (\s_2)\}
= \mp\d^{\prime} (\s_1 - \s_2 ) (T\sb \pm (\s_1) +
T\sb \pm (\s_2)) \eqno(13)$$
under the Poisson brackets (5).
The currents $J_{A\a}$ satisfy the Kac-Moody algebra
$$ \{J\sb{\pm \a} (\s_1),J\sb{\pm \b} (\s_2)\}
= f\sb{\a\b}\sp{\g} J\sb{\pm \g} (\s_1) \d (\s_1 - \s_2)
\pm 2\k  \g_{\a\b} \d^{\prime} (\s_1 - \s_2 )\quad.\eqno(14)$$
As usual, the plus and minus currents have
vanishing Possion brackets.

Formulas (11) and (12) are the basis for building a canonical
representation of a $W$ algebra, since we can write
$$ W\sb{A s} = \frac1s d\sp{\a\sb 1 \cdots \a\sb s}J\sb{A \a\sb 1}
\cdots J\sb{A \a\sb s} \quad (s=2,...,N)\quad. \eqno(15)$$
The coeficients $d\sp{\a_1 ... \a_s}$ will be determined from the
requirement that the Poisson bracket algebra of $W$'s closes
(or equivalently, $W$'s are first class constraints).
The results of [7,11,12] imply that the general relation (15)
can be simplified to
$$ W\sb{A s} = {1\over 2\k s} Tr (J\sb{A}\sp s )\quad, \eqno(16)$$
where $J_A = J_A^\a t_\a $.
In the case of the $W_3$ algebra we have [12]
$$  W_{A3} = {1\over 6\k} Tr (J_A^3) \quad,\eqno(17)$$
so that $d_{\a\b\g} = {1\over 4\k} Tr (t_\a \{ t_\b , t_\g\}) $. One can
check that $T$ and $W$ given by (11) and (17) form a classical $W_3$ algebra
$$\li{\{T\sb \pm (\s_1),T\sb \pm (\s_2)\}
&= \mp\d^{\prime} (\s_1 - \s_2) (T\sb \pm (\s_1) +
T\sb \pm (\s_2)) \cr
\{T\sb \pm (\s_1),W\sb \pm (\s_2)\}
&= \mp\d^{\prime} (\s_1 - \s_2) (W\sb \pm (\s_1) +
2W\sb \pm (\s_2)) \cr
\{W\sb \pm (\s_1),W\sb \pm (\s_2)\}
&= \mp 2c\d^{\prime} (\s_1 - \s_2) (T\sp 2\sb \pm (\s_1) +
T^2\sb \pm (\s_2)) \quad,&(18)\cr}$$
and all other Poisson brackets are zero. Here $W=W\sb 3$, while
$c=c_1 - \frac1n$, where $c_1$ is a constant defined by the relation
$$ Tr(J^4) = c_1 (Tr J^2)^2 \quad.\eqno(19)$$
The relation (19) is valid for $G=A_l , B_l , C_l$, $l\le 2$, since for the
groups of rank $l>2$, $Tr(J^4)$ is an independent Casimir invariant [12].

The 2d diffeomorphism invariance requires
$H\sb 0 =0$. Otherwise, the wavefunctional $\Psi[\f]$ would depend explicitely
on the unphysical
parameter $\t$, since $i\der{\t}\Psi = \hat{H}\sb 0 \Psi $. Then according to
the Eq. (3) the gauge invariant action is simply
$$ S\sb N = \int_{\t_1}^{\t_2} d\t \int_0^{2\p} d\s
\left( \p\sb a \dt{\f}\sp a - h\sp{A} T\sb{A}
- \su_{s=3}^{N} b\sp{A}\sb s W\sb{A s}\right) \quad,\eqno(20)$$
where $h$ and $b$ are the
lagrange multipliers, which are also the gauge fields corresponding to
the $W$-symmetries. The gauge transformation laws can be determined from
the Eq. (4). In the $W\sb 3 $ case they become
$$\li{\d \p\sb a =& \left( \e^A {\g^{\a\b} \over 2\k} +
\x^A d^{\a\b\g}J_{A\g}\right) J_{A\a} {\pa J_{A\b}\over \pa \f^a} \cr
&+ \k \left[\left( \e^A {H\sp{bc} \over 2\k} +\x^A D\sb A\sp{bcd} J\sb{Ad}
\right)J\sb{Ab} ( 2\ct_{ca}+(-1)^A H_{ca} ) \right]^{\prime}
,&(21.a)\cr
\d \f\sp a =& {\e\sp A \over 2\k} J\sb{A}\sp{a}
+\x\sp A D\sb{A}\sp{a}\sb{bc}J\sb{A}\sp{b} J\sb{A}\sp{c}
\quad,&(21.b)\cr
\d h\sp A =& \dt{\e}\sp A -(-1)^A [ h\sp A (\e\sp A)^{\prime} -
(h\sp A )^{\prime}\e\sp A] + 2c(-1)^A [\x\sp A (b\sp A)^{\prime} -
(\x\sp A)^{\prime} b\sp A ] T\sb A \quad,&(21.c)\cr
\d b\sp A =& \dt{\x}\sp A +(-1)^A \left[ 2(h\sp A)^{\prime}\x\sp A
- h\sp A (\x\sp A )^{\prime} - 2 b\sp A (\e\sp A)^{\prime}
+ (b\sp A )^{\prime}\e\sp A\right] \quad,&(21.d)\cr}$$
where $\e^A$ are the parameters of
the $T\sb A$ transformations, $\x^A$ are the parameters of the
$W\sb A$ transformations, while
$$ D_{Aabc}(\f) = d_{\a\b\g}E_{Aa}^\a (\f) E_{Ab}^\b (\f) E_{Ac}^\g (\f)
\quad,\quad
J_{Aa} = E_{Aa}^\a J_{A\a} \quad.\eqno(22)$$
In all equations we use Einstein's summation convention, i.e. summation is
performed only if the up and down index are the same.

In order to find a geometrical interpretation of the action (20) we need
to know its second order form.
It can be obtained by replacing the momenta $\p\sb a$ in (20) by their
expressions in terms of $\f^a$. These expressions can be obtained
from the equation of motion
$${\d S_N \over \d \p\sb a} = 0\quad.\eqno(23)$$
In the $W\sb 3$ case one gets
$$ \dt{\f}\sp a + {h\sp A\over 2\k} J_A^a +
b\sp A D\sb{A}\sp{a}\sb{bc} J_A^b J_A^c  = 0
\quad.\eqno(24)$$
This is a quadratic equation in  $\p_a$, and therefore the second order
form of the Lagrangian density
will be a non-polynomial function of $\pa_\m \f$,
$h$ and $b$, which can be written
as an infinite power series in those variables. There is a complete analogy
with the flat background (or Abelian $G$) case [9], where
$$ J_{\pm}^a = -H^{ab}\p_b - 2\k \ct\sp{a}\sb{b}\f^{\prime b}
\mp \k\f^{\prime a} \to - \p_a \mp \k\f^{\prime a} \quad.
\eqno(25)$$
Note that in the case of an arbitrary background $H_{ab}$, the expression
(25) for $J$ (or equivalently Eq. (12)) is not useful for building the
generators of a $W$ algebra since then the $J$'s do not
satisfy the Poisson bracket Kac-Moody algebra (14).

As a preparation for the $W_3$ case, we first study the second order action
obtained from the first order action (10) in the $W_2$ case. One can show that
after the elimination of the momenta one obtains the covariant form of the
WZNW action (6), after the following identifications
$$ \tilde{g}\sp{00} = {2\over h\sp + + h\sp -} \quad,\quad
\tilde{g}\sp{01} = {h\sp - - h\sp + \over h\sp + + h\sp -} \quad,\quad
\tilde{g}\sp{11} = - {2h\sp + h\sp -\over h\sp + + h\sp -} \quad,\eqno(26)$$
where $\tilde{g}\sp{\m\n} = \sqrt{-g}g\sp{\m\n}$. The covariant form of the
2d diffeomorphism transformations can be obtained from the Eq. (21.b),
by rewritting it as
$$ \d\f\sp a = {\e\sp A \over 2\k} J\sb{A}\sp{a} =
-{\e\sp A\over\sqrt{ h^+ + h^- }}\tilde{e}\sb A\sp \m \pa\sb \m
\f\sp a = \e\sp \m \pa\sb \m \f\sp a \quad,\eqno(27)$$
where
$$ \tilde{e}\sb{A}\sp \m = {1\over \sqrt{h\sp + + h\sp -}}
\pmatrix{1 &h\sp - \cr 1 &-h\sp +\cr} \quad.\eqno(28)$$
Eq. (21.c) can be rewritten as
$$\d\tilde{g}\sp{\m\n} = -\pa\sb \r (\e\sp \r \tilde{g}\sp{\m\n}) +
\pa\sb \r \e\sp{(\m|}\tilde{g}\sp{|\n)\r} \quad,\eqno(29)$$
which is the
diffeomorphism transformation of a densitized metric
generated by the parameter $\e\sp \m$. The metric $g\sp{\m\n}$ can be written
as
$$g\sp{\m\n}= {1\over \sqrt{-g}(h\sp + + h\sp - )}
\pmatrix{2 &h\sp - - h\sp +\cr h\sp - - h\sp + &-2h\sp + h\sp - \cr}
= e\sb +\sp{(\m|} e\sb -\sp{|\n)} \quad, \eqno(30)$$
where $e\sb A\sp \m =(-g)^{-{1 \over 4}} \tilde{e}\sb A\sp \m$ are
the zweibeins.
Note that $\sqrt{-g}$ is undetermined, because
the action (6) is independent of $\sqrt{-g}$ due to the Weyl
symmetry
$$ \d g\sp{\m\n} = \o g\sp{\m\n}\quad. \eqno(31) $$
Also note that the relations  (26,28,30) are essentialy the same as in the
flat background case [9].

In the $W\sb 3$ case we have from the Eq. (24)
$$\li{ \p_a + 2\k\ct_{ab}\f^{\prime b} &=
\k H_{ab}\,\tilde{g}^{0\m}\pa_\m \f^b + \d\P_a \cr
\d\P_a &= {b\sp A \over h^+ + h^-}D\sb{Aabc}J_A^b J_A^c
 \quad, &(32)\cr}$$
where $\tilde{g}^{0\m}$ is given by the Eq. (26).
Then the action (20) takes the following form
$$ S\sb 3 = \int d^2 \s \left( \cl_{2}
-{h\sp + + h\sp - \over 4 \k}\d\P^a \d\P_a
-{ b\sp{A}\over 3}D\sb{Aabc}J_A^a J_A^b J_A^c \right)
\quad,\eqno(33)$$
where $\cl_2$ is the Lagrangian density of the WZWN action (6).
Note that the Eq. (24) can be rewritten as
$$ J\sb{A}^a =-{2\k\over \sqrt{h^+ + h^-}}\pa\sb A \f\sp a -
{b\sp A \over h\sp + + h\sp -}D\sb{A}\sp{a}\sb{bc}J_A^b J_A^c
\quad.\eqno(34)$$
Eq. (34) can be used to obtain a power series expansion of $J_A$ in terms
of $\pa\sb \pm\f$, $h$ and $b$, which can be inserted into Eq. (33) to give
the  corresponding power series expansion of the action. Up to the first order
in $b$ the Lagrange desity can be written as
$$ \cl_3 = \cl_{2} - b\sp{ABC} D\sb{+abc}\pa\sb{A}\f\sp a
\pa\sb{B}\f\sp b \pa\sb C\f\sp c  + O(b^2)\quad,
\eqno(35)$$
where the only nonzero components of $b\sp{ABC}$ are
$$ b\sp{\pm\pm\pm} =\pm {4\k^3\over 3 }{b\sp \pm \over ( h\sp +
+ h\sp -)^{3 \over 2} }\quad,\eqno(36)$$
and we used the property $D_{+abc} = -D_{-abc}$.

It is clear that the above procedure will give the following form of
the second order covariant Lagrange density
$$\li{\cl =& \cl_{2} +  \tilde{b}\sp{\m\n\r}D\sb{abc}(\f)
\pa\sb \m \f\sp a \pa\sb \n \f\sp b \pa\sb \r \f\sp c \cr
&+ \tilde{c}\sp{\m\n\r\s}D\sb{ab}\sp{e}(\f)D\sb{ecd}(\f)
\pa\sb \m \f\sp a \pa\sb \n \f\sp b \pa\sb \r \f\sp c \pa\sb \s \f\sp d
+ \cdots \quad,&(37)\cr}$$
which for an Abelian $G$ reduces to the flat-space case [9].
The objects $\tilde{g}$, $\tilde{b}$, $\tilde{c}$,
... , must transform
as tensor densities since the action is invariant under the infinitesimal
diffeomorphisms
$$ \d \f^a = {\e^A\over 2\k} J_A^a = {\e}^\m \pa_\m \f^a \quad,\quad
{\e}^\m = f^\m (\e^\pm , h^\pm ,
b^\pm,\f^a, \pa_\m \f^a ) \quad.\eqno(38)$$

Besides the diffeomorphism invariance, the generalized Weyl symmetry [1]
is also
obscured. Heuristically it is there by construction, since we used only four
independent gauge fields $h\sp \pm$ and $B\sp \pm$. The fields $\tilde{g}$,
$\tilde{b}$, $\tilde{c}$, ... in the Eq. (37) are functions of $h$ and $b$,
and one can check order by order in $\pa\f$ that
$$ \tilde{g}\sb{\m\n}\tilde{b}\sp{\m\n\r} = 0 \quad,\quad
   \tilde{c}\sp{\m\n\r\s} = \tilde{g}\sb{\t\e}\tilde{b}\sp{\m\n\t}
\tilde{b}\sp{\e\r\s} \quad, \eqno(39)$$
and so on, which is the covariant form of the generalized Weyl symmetry.

In conclussion we can say that the propagation of the bosonic $W_3$ string
on a curved background is described by a non-polynomial action
whose Lagrange density is given by the Eq. (37). This action
is of the similar form as the action in the flat background case [9],
and the only difference is that the $d_{\a\b\g}$
coeficients become functions of the fields $\f^a$ via the Eq. (22). It remains
to be explored how to generalize the transformation laws (21.a-b) to the case
of
an arbitrary background $H_{ab}(\f)$.

Note that for a realistic W-string theory the group $G$ has to be
non-compact. When $G$ is compact, the space-time metric $H_{ab}$ is of the
Euclidian
signature, and moreover, there are no propagating degrees of freedom
classically, since the $T_A$ constraints imply
$$ J^a_{\pm} = 0 \to\quad \p_a = 0 \,\, , \,\, \f^{\prime a} = 0 \quad.
\eqno(40)$$
In the quantum case one can get propagating degrees of freedom due to the
anomalies which will appear in the $W$ algebra
(see [13,14] for the $W_2$ case).
Still, the compact $G$ construction can be relevant if the spacetime manifold
is of the type $ M^d \times G $
where $M^d$ is the $d$-dimensional Minkowski spacetime. However, one has to
keep
in mind that due to the nonlinearity of the $W$ algebra
(except in the $W_2$ and $W_\infty$ case) one cannot construct
a representation for $M^d \times G$
by adding the representations for  $M^d$ and $G$.

\refs
\Item{[1]}C.M. Hull, Classical and quantum W-gravity, in Strings and
Symmetries 1991, eds. N. Berkovits et al., World Scientific, Singapore (1992)
\Item{[2]}A.B. Zamolodchikov, Teor. Mat. Fiz. 65 (1985) 1205\\
          V.A. Fateev and A.B. Zamolodchikov, \NP 280 (1987) 644\\
          V.A. Fateev and S. Lukyanov, Int. J. Mod. Phys. A3 (1988) 507
\Item{[3]}C.M. Hull, \PL 240B (1990) 110
\Item{[4]}K. Schoutens, A. Sevrin and P. van Nieuwenhuizen, \PL 243B (1990) 245
\Item{[5]}E. Bergshoeff, C.N. Pope, L.J. Romans, E. Sezgin, X. Shen and K.S.
          Stelle, \PL 243B (1990) 350
\Item{[6]}C.M. Hull, \NP 353 (1991) 707
\Item{[7]}C.M. Hull, \PL 259B (1991) 68
\Item{[8]}A. Mikovi\'c, \PL 260B (1991) 75
\Item{[9]}A. Mikovi\'c, \PL 278B (1991) 51
\Item{[10]}G.M. Sotkov, M. Stanishkov and C.-J. Zhu, \NP 356 (1991) 245, 439
\Item{[11]}J. Thierry-Mieg, Lectures at the Cargese School on Nonperturbative
            Quantum Field Theory (1987)\\
            F. Bais, P. Bouwknegt, M. Surridge and K. Schoutens, \NP 304 (1988)
            348, 371
\Item{[12]}J. Balog, L. Feher, P. Forgacs, L. O'Raifertaigh and A.
Wipf, Ann. Phys. 203 (1990) 76
\Item{[13]}D. Gepner and E. Witten, \NP 278 (1986) 493
\Item{[14]}G. Felder, K. Gawedzcki and A. Kupianen, \CMP 117 (1988) 127

\end{document}